\documentclass[10pt]{article}

\usepackage{graphics}
\usepackage{graphicx}

\usepackage[a4paper, left=35mm,right=35mm,top=34mm,bottom=34mm]{geometry}
\usepackage[utf8]{inputenc}
\usepackage[T1]{fontenc}
\usepackage[english]{babel}

\usepackage{enumerate}
\usepackage{graphicx}
\usepackage{hyperref}
\hypersetup{
	colorlinks=true,
	linkcolor=blue,
	filecolor=magenta,      
	urlcolor=cyan,
}
\usepackage{listings}
\usepackage{color}

\definecolor{dkgreen}{rgb}{0,0.6,0}
\definecolor{gray}{rgb}{0.5,0.5,0.5}
\definecolor{mauve}{rgb}{0.58,0,0.82}

\usepackage{mathtools,amsthm,amssymb,amsfonts}
\usepackage{algorithm}
\usepackage{algorithmic}
\makeatother
\theoremstyle{plain}

\theoremstyle{definition}

\theoremstyle{remark}

\usepackage{caption} 
\usepackage{fancyhdr}
\usepackage{subcaption}
\usepackage{pgfplots}\pgfplotsset{compat=newest}

\usepackage{float}
\usepackage{pythonhighlight}
\lstnewenvironment{mypython}[1][]{\lstset{style=mypython,#1}}{}

\usepackage{mathtools}
\DeclarePairedDelimiterX\SGprod[2]{\langle}{\rangle_G}{#1,#2}
\DeclarePairedDelimiterX\SGnorm[1]{\vvvert}{\vvvert}{#1}

\usepackage[normalem]{ulem}

\author{
  {\normalsize Ahmed El Kerim}\thanks{Laboratory of Mathematics in Interaction with Computer Science, CentraleSup\'elec, Paris-Saclay University, Gif-sur-Yvette, France.}
  \and
  {\normalsize Pierre Gosselet}\thanks{Lamcube, University of Lille, CNRS, Centrale Lille, Lille, France}
  \and
  {\normalsize Fr\'ed\'eric Magoul\`es}\thanks{Laboratory of Mathematics in Interaction with Computer Science, CentraleSup\'elec, Paris-Saclay University, Gif-sur-Yvette, France.}
}
\title{Enhancing the Global/Local Coupling Method: An Asynchronous Parallel Framework}

\begin{document}
\maketitle
\thispagestyle{fancy}

\begin{abstract}
\noindent A novel approach is being developed to introduce a parallel asynchronous implementation of non-intrusive global-local coupling. This study examines scenarios involving numerous patches, including those covering the entire structure. By leveraging asynchronous, the method aims to minimize reliance on communication, handle failures effectively, and address load imbalances. Detailed insights into the methodology are presented, accompanied by a demonstration of its performance through an academic case study.
\end{abstract}


\section{Introduction}
In the industry, the submodeling method \cite{kelley1982, ransom1992computational, cormier_Aggressive_1999} has been widely employed for studying local complex behaviors in simulated structures. However, despite its non-intrusive nature, this approach often results in significant errors, which are evident as imbalances at the interfaces of the submodel patches in the reference model.

To overcome these limitations and enhance the accuracy of the submodeling method, the global/local coupling technique \cite{gendre2009non} has been developed. This technique adopts an iterative approach to refine the submodeling outcomes and achieve greater precision. The global/local coupling method can be classified as a domain decomposition technique, specifically a Schwarz decomposition \cite{hecht_2009_nzs, blanchard.2018.1, duval2014non, ELKERIM2023115910}. These methods are well-suited for parallel computation, and recent studies \cite{magoules2018asynchronous, Garay2022SynchronousAA, Glusa, Gbikpi-Benissan2022} have demonstrated their compatibility with asynchronous iterations. Asynchronous iterations provide enhanced tolerance to network latencies, load imbalances, and highly heterogeneous computing architectures.
\vspace*{2.5mm}
In the paper \cite{ELKERIM2023115910}, the first asynchronous version of the global/local coupling method is introduced for linear elliptic problems. Building upon this work, the current paper briefly presents the coupling method and its asynchronous variant. Furthermore, it discusses the implementation aspects and showcases novel numerical results obtained through this approach.

\section{Principle of the method}
\begin{figure}[h!!]
	\centering
	\begin{subfigure}{0.32\textwidth}
		\centering
		\includegraphics[width=\textwidth]{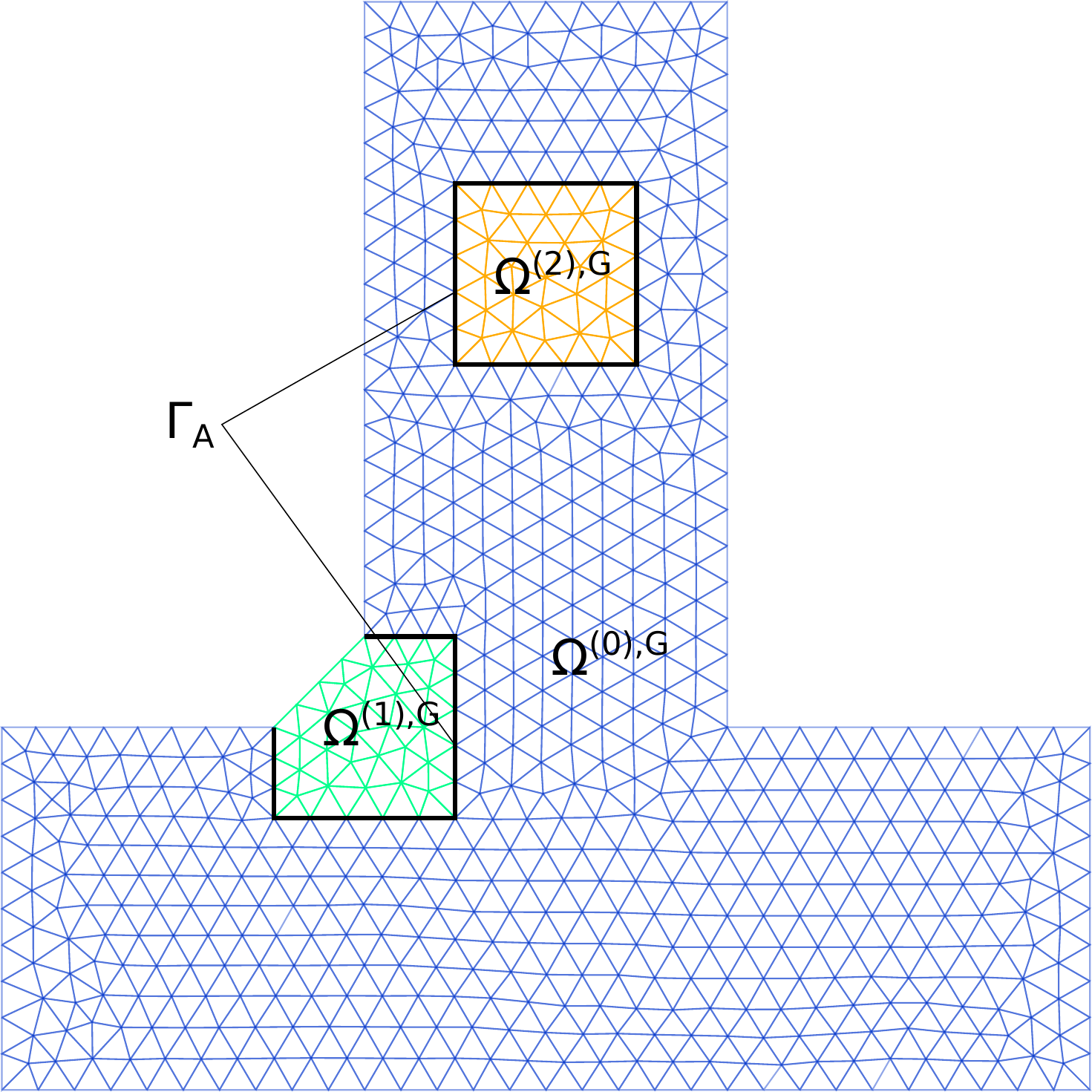}
		\caption{Global problem}
		\label{Global model}
	\end{subfigure}
	\begin{subfigure}{0.32\textwidth}
		\centering
		\includegraphics[width=0.5\textwidth]{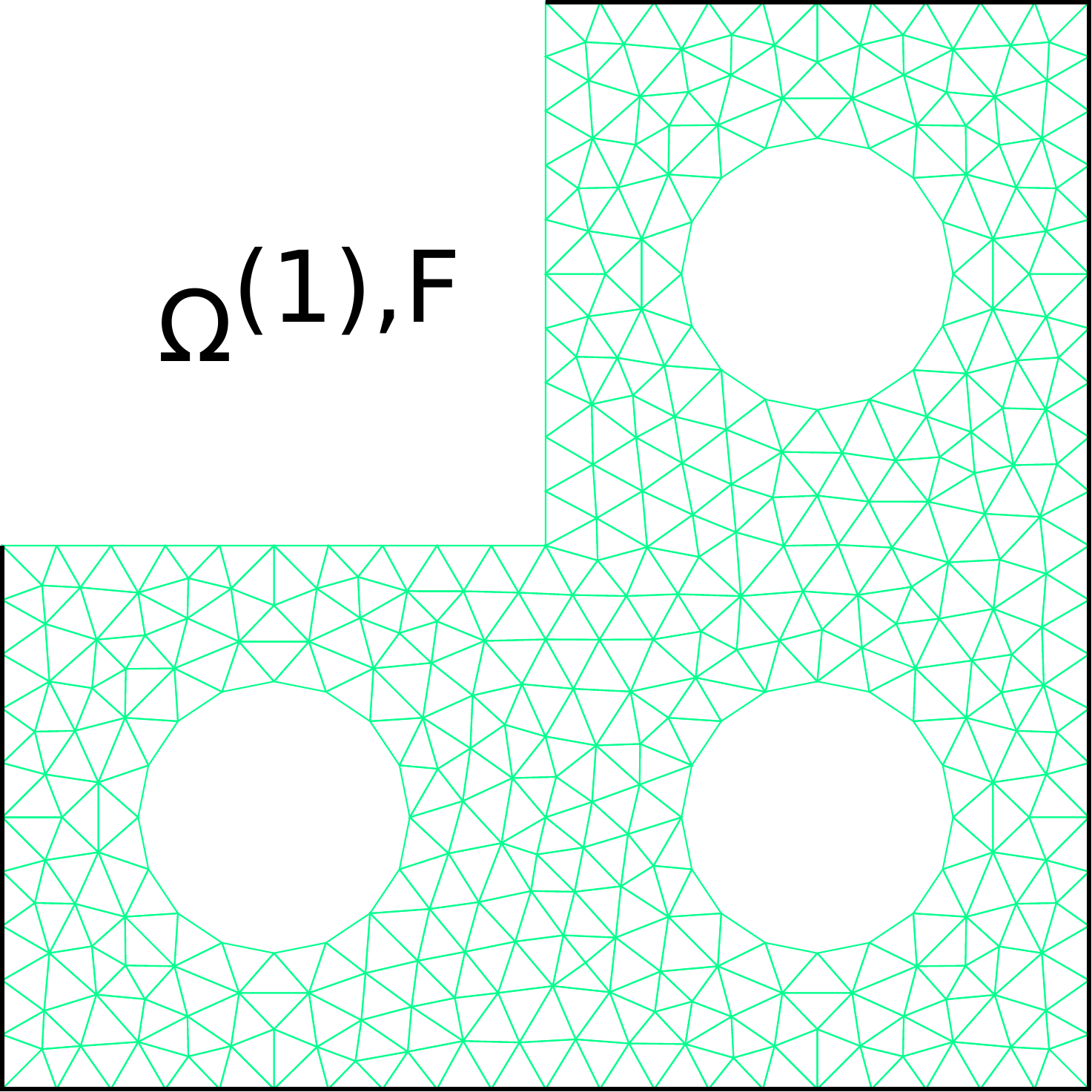}
		\includegraphics[width=0.5\textwidth]{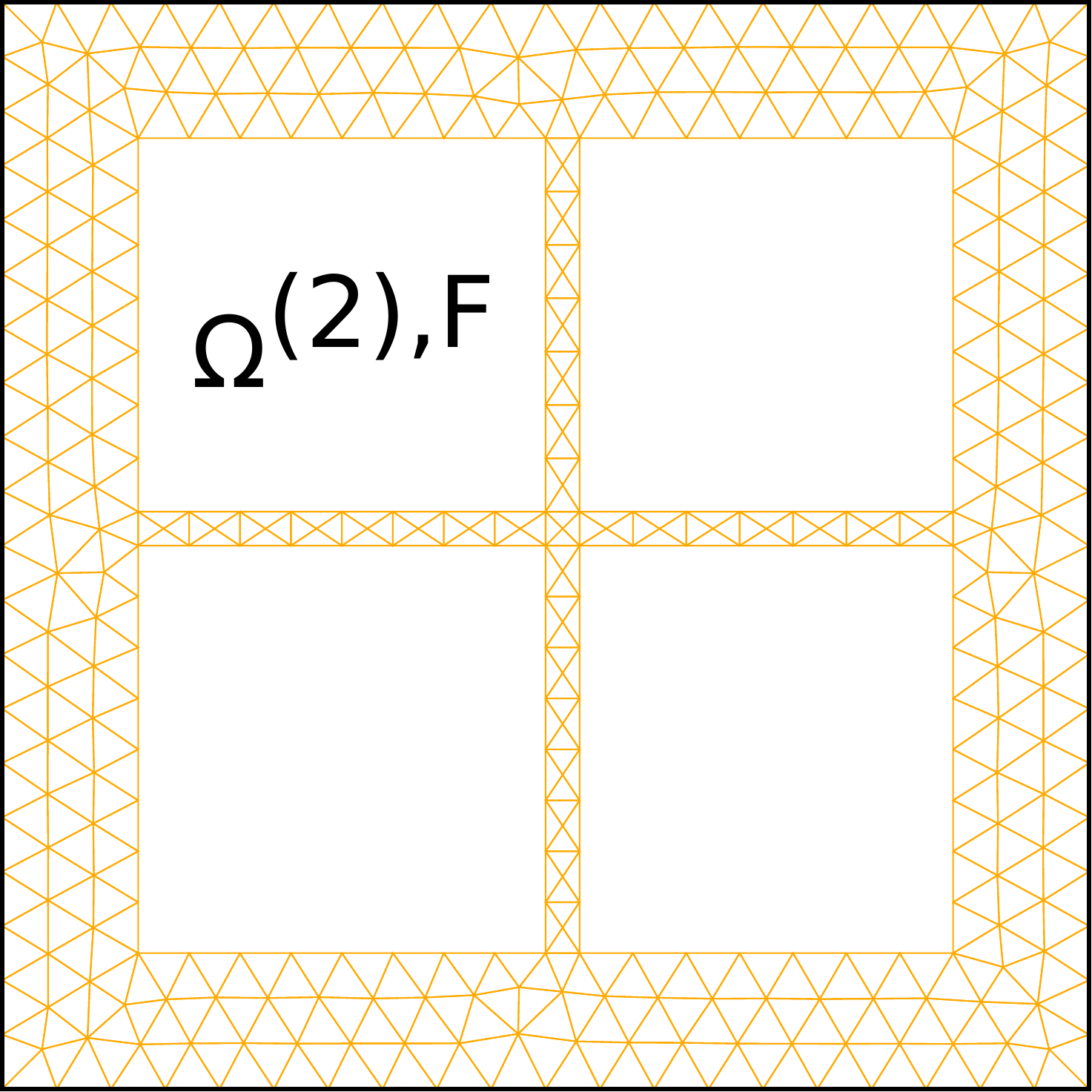}
		\caption{Refined zones of interest}
		\label{Zones of interest}
	\end{subfigure}
	\begin{subfigure}{0.32\textwidth}
		\centering
		\includegraphics[width=\textwidth]{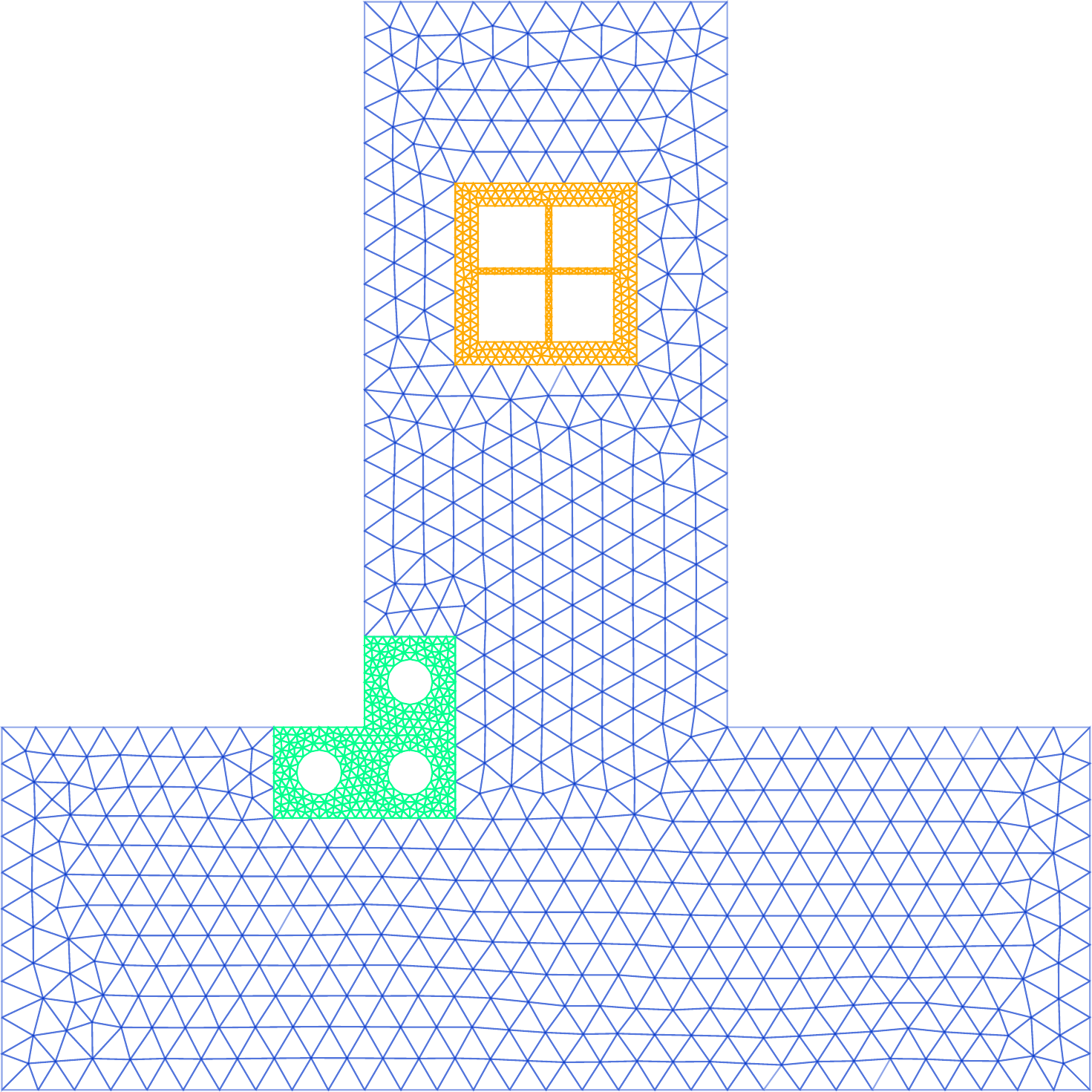}
		\caption{Reference problem}
		\label{ref_model}
	\end{subfigure}
	\caption{Coarse and refined meshes}
	\label{2D}
\end{figure}

To illustrate the study of local complex behaviors in simulated structures, we consider an approximation of a 2D turbine blade, as shown in Figure \ref{2D}. The model comprises three distinct zones: a global zone (depicted in blue), and two zones of interest (highlighted in yellow and green) that exhibit specific and intricate geometries.

In Figure \ref{Global model}, we present the coarse-scale representation of the 2D turbine blade, referred to as the global model. The coarse mesh used in this model cannot accurately capture the behaviors occurring in the zones of interest. Therefore, highly refined meshes are required to simulate these zones effectively, as shown in Figure \ref{Zones of interest}. The refined zones of interest, denoted as $\Omega^{1,F}$ and $\Omega^{2,F}$, are where geometrical modifications or nonlinear behaviors, such as plasticity, viscoplasticity, fatigue, and cracking, are of particular interest.

\begin{figure}[h!!!]
	\includegraphics[width=0.98\textwidth]{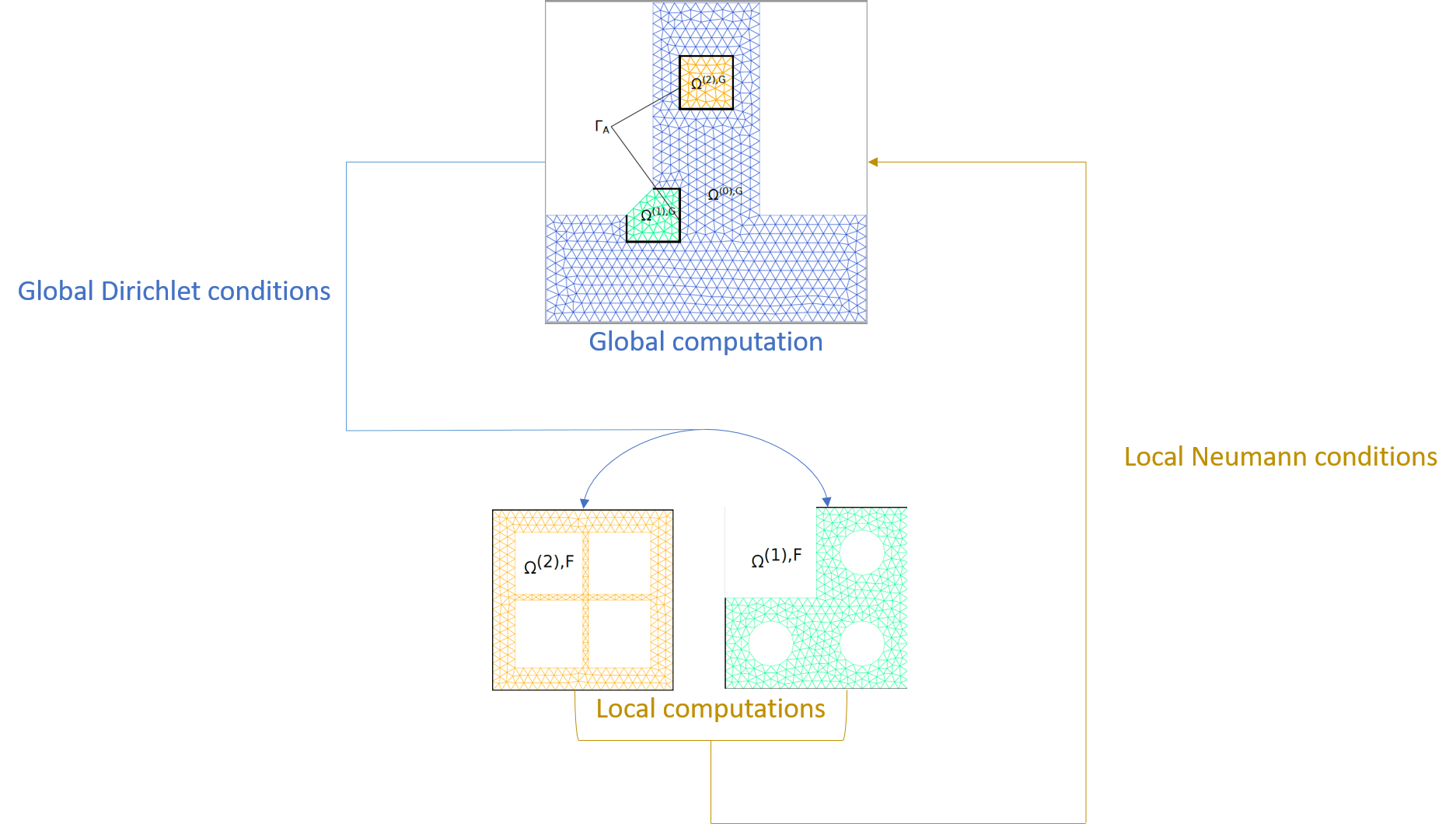}
	\caption{Global Local non-invasive coupling iteration}
	\label{fig:GL_scenario}	
\end{figure}

The classic scenario of the global/local coupling method is illustrated in Figure \ref{fig:GL_scenario}. Initially, a linear global coarse model represents the overall structure (Figure \ref{Global model}). Subsequently, specific zones of interest $\Omega^s_G$ $(s>0)$ are identified based on criteria or the need for more detailed modeling. In the example shown, adapted meshes and geometrical details are introduced in the fine modeling of these zones, and modifications to material laws may also be incorporated. The fine computations are performed in parallel on the patches, utilizing the global solution as a Dirichlet boundary condition. Although the fine and global subdomains may differ ($s>0$), their interface $\Gamma^s$ remains consistent, defined as $\Gamma^s=\Omega\cap\partial\Omega^{s,F}=\Omega\cap\partial\Omega^{s,G}$.
The error in the flux balance between the global zone $\Omega^0$ (not covered by patches) and the fine models is quantified as a lack of equilibrium. This error, known as the residual, is incorporated into the global model as an immersed Neumann condition on the interface. The corrected global model, accounting for this interface load, is then solved, followed by the solution of local Dirichlet problems. This iterative process continues until convergence is achieved.

The global/local coupling method can be viewed as a simple iterative technique, specifically a Richardson iteration in its simpler form. Its objective is to obtain the reference solution (Figure \ref{ref_model}) by utilizing computations performed on the global and fine models, without the need for creating a potentially cumbersome reference model or making significant interventions in the models and software.

It is worth noting that if the computation is halted after the first iteration without sending back the nodal reactions to the global model, a submodeling method is performed.

\section{Asynchronous coupling iterations}
\begin{figure}[H]\centering
	\begin{subfigure}{.45\textwidth}\centering
		\includegraphics[width=0.8\textwidth]{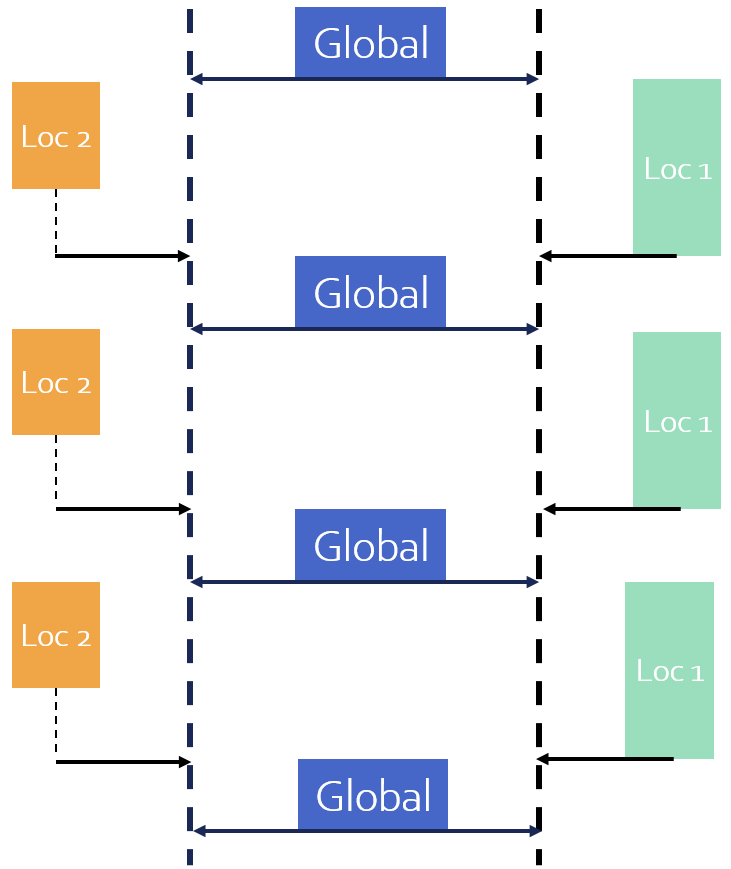}
		\caption{Synchronous iteration}
		\label{Synchronous model}
	\end{subfigure}
	\includegraphics[width=0.07\textwidth]{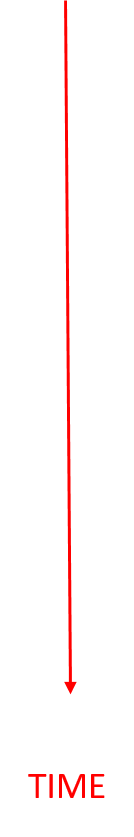}
	\begin{subfigure}{.45\textwidth}\centering
		\includegraphics[width=0.8\textwidth]{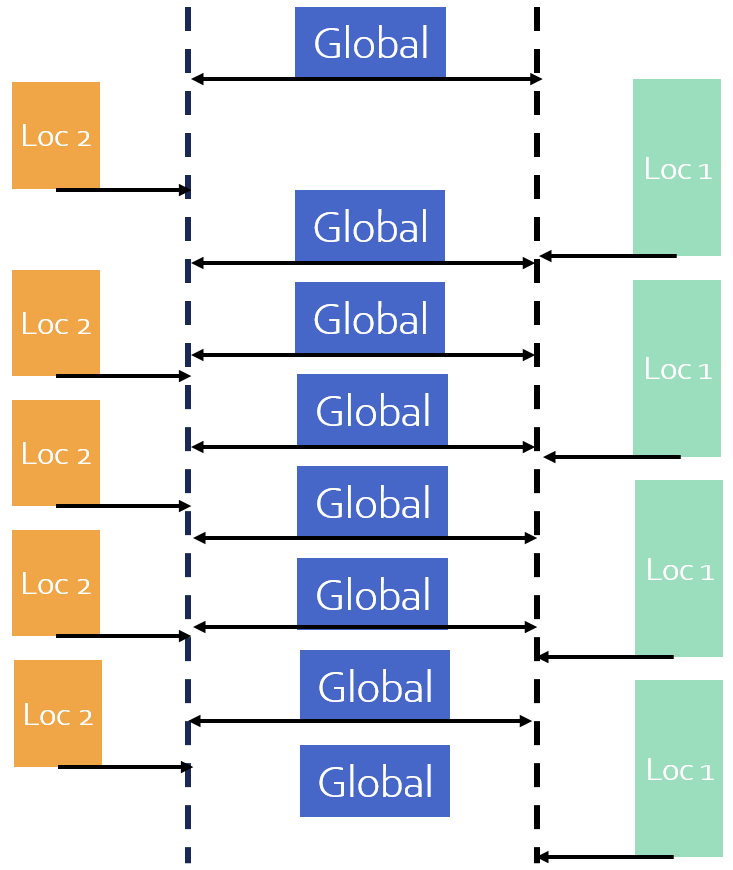}
		\caption{Asynchronous iteration}
		\label{Asynchronous with wait}
	\end{subfigure}
	\caption{Time course of the Global/Local coupling in the case of two patches}\label{fig:timecourse}
\end{figure}

The Global/Local coupling method has been recognized as a robust and non-invasive technique. However, its applicability to large-scale simulations and modern parallel computing architectures is limited due to its alternating nature, as discussed in \cite{ELKERIM2023115910}. This alternating approach, illustrated in Figure \ref{Synchronous model}, introduces waiting and inactivity times on both sides, leading to reduced performance, especially when faced with load imbalances, communication delays, or machine failures.

To address these challenges and improve the efficiency of the Global/Local coupling method, an asynchronous parallel version has been proposed. This approach allows each processor to work independently and at its own pace, utilizing the most recent available data. The time sequence depicted in Figure \ref{Asynchronous with wait} illustrates the asynchronous approach, where processors only wait when they have no new data to process.

By adopting asynchronous iterations, the method gains several advantages. It becomes more resilient to network latencies, load imbalances, and heterogeneous computing architectures. Furthermore, by reducing waiting and inactivity times, the overall performance and efficiency of the Global/Local coupling method are enhanced, making it well-suited for high-performance computing environments.

The asynchronous approach enables processors to make progress independently and asynchronously, effectively utilizing computational resources and minimizing potential bottlenecks. This adaptability to modern parallel computing architectures and the ability to handle large-scale simulations make the asynchronous parallel version of the Global/Local coupling method a favorable choice.

By embracing the benefits of the asynchronous parallel version, we can overcome the limitations of the synchronous approach and achieve improved performance and scalability in high-performance computing environments. This advancement opens up new possibilities for effectively applying the Global/Local coupling method in a wide range of complex simulations and analyses.

\section{Implementation}
RDMA (Remote Direct Memory Access) has emerged as a promising approach for implementing asynchronous communication protocols. Unlike traditional two-sided communication based on MPI, RDMA utilizes one-sided communication, which has shown significant efficiency and adaptability in recent research. Works in \cite{Yamazaki.19} and \cite{Glusa} highlight the effectiveness of one-sided communication, also known as MPI-RDMA, in achieving efficient asynchronous communication.

To implement RDMA-based asynchronous communication, a memory region called a window is created. This window allows direct access to data on remote machines without involving the target machine. PUT and GET operations can be performed on the window, enabling efficient data transfer and retrieval. This approach eliminates the need to interrupt computations for send or receive operations, making it highly suitable for asynchronous calculations.

The experiments reveal that network-specific factors, MPI version variations, and the inherent asynchronicity of communication influence the overall performance of RDMA-based asynchronous protocols.

Synchronization is an essential aspect of RDMA communication. Two synchronization techniques are commonly used: active synchronization and passive synchronization. Active synchronization involves performing collective operations using \textbf{MPI\_WIN\_Fence()} to synchronize all processes before transitioning between iterations. In contrast, passive synchronization is employed in asynchronous scenarios. Processors individually synchronize by opening an epoch with \textbf{MPI\_win\_lock}, executing desired \textbf{PUT} and \textbf{GET} operations, and then closing the epoch with \textbf{MPI\_win\_Unlock}. The \textbf{MPI\_win\_Flush} operation ensures the completion of send operations within the epoch.

Overall, RDMA-based asynchronous communication, leveraging the advantages of one-sided communication and appropriate synchronization techniques, offers a promising approach for efficient and scalable implementations in various MPI versions and network configurations.
\section{Numerical results}
\subsection{Codes}
In our Python code implementation, we utilize various tools and software to facilitate different aspects of the study. Firstly, we employ GMSH \cite{GMSH} for generating geometries and meshes for the cases under investigation. For the finite element approximation, we rely on the Getfem library \cite{GetFEM}. To enable parallel computing, we utilize the mpi4py library \cite{mpi4py}.

The study was conducted using the cluster at the LMPS simulation center, utilizing multiple workstations interconnected via an Ethernet network. The machines employed in the cluster exhibit heterogeneity, featuring four different generations of CPUs: Intel(R) Xeon(R) CPU E5-1660 v3 (Haswell) @ 3.00GHz, Intel(R) Xeon(R) CPU E5-2630 v4 (Broadwell) @ 2.20GHz, Intel(R) Xeon(R) Silver 4116 CPU (Skylake) @ 2.10GHz, and Intel(R) Xeon(R) W-2255 CPU (Cascade Lake) @ 3.70GHz.

\subsection{Studied case}
\begin{figure}[ht]
	\null\hfill
	\begin{subfigure}[b]{0.63\linewidth}	\centering 
			\includegraphics[width=0.9\textwidth]{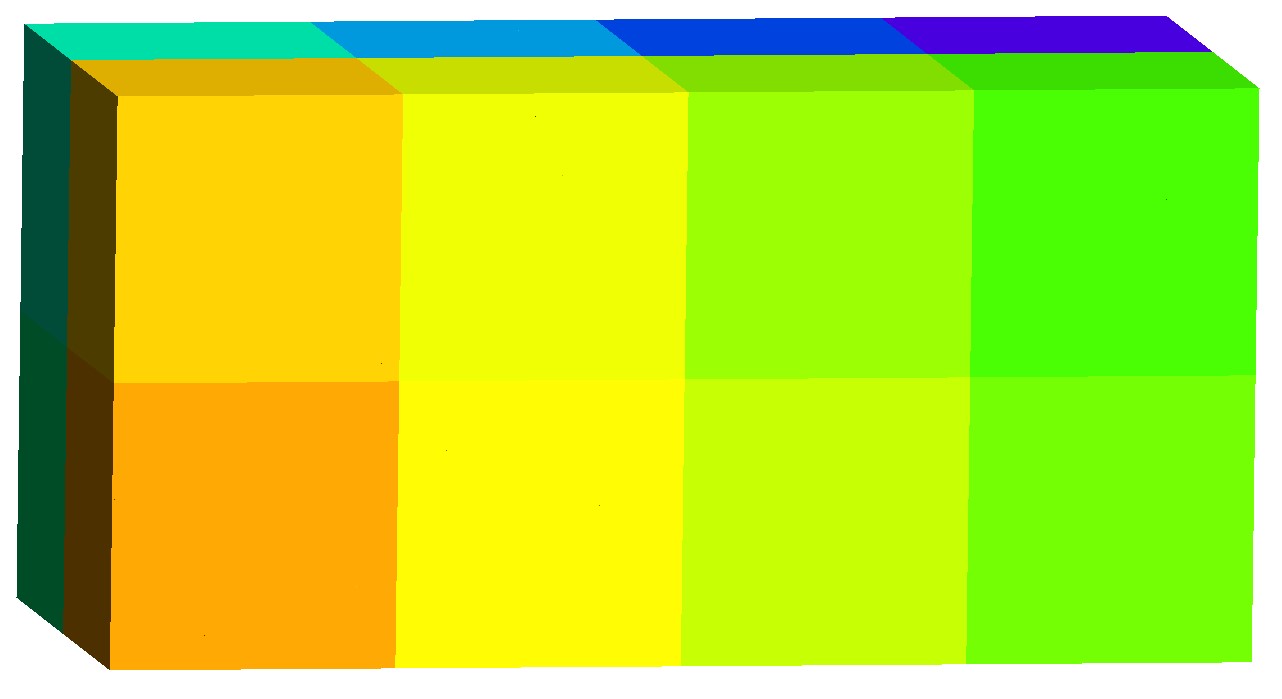}
			\caption{16 patches}
			\label{16_cube}
	\end{subfigure}
	\hfill		
	\begin{subfigure}[b]{0.35\linewidth}	\centering 
			\includegraphics[width=.9\textwidth]{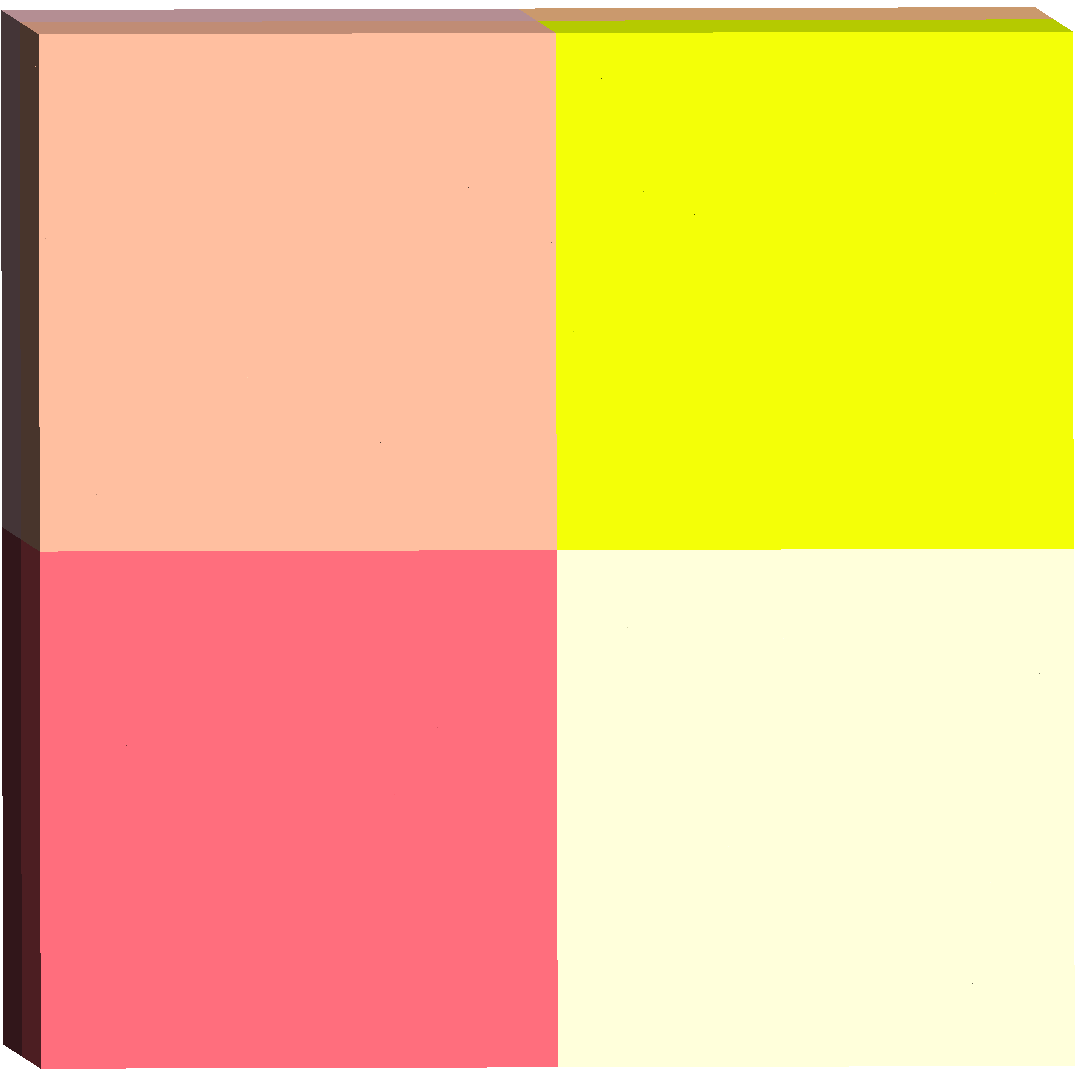}
			\caption{8 patches}
			\label{8_cube}
	\end{subfigure}
	\hfill\null
\end{figure}

The aim of the study is to generate a flexible number of patches, specifically non-overlapping contiguous patches that cover the entire global model. To achieve this, a cuboid domain is chosen, with cuboid patches applied. An example of an 8-patch configuration can be seen in Figure~\ref{8_cube}, while Figure~\ref{16_cube} illustrates a 16-patch scenario.

\begin{figure}[ht]\null\hfill
	\begin{subfigure}{.35\textwidth}
		\centering
		\includegraphics[width=0.9\textwidth]{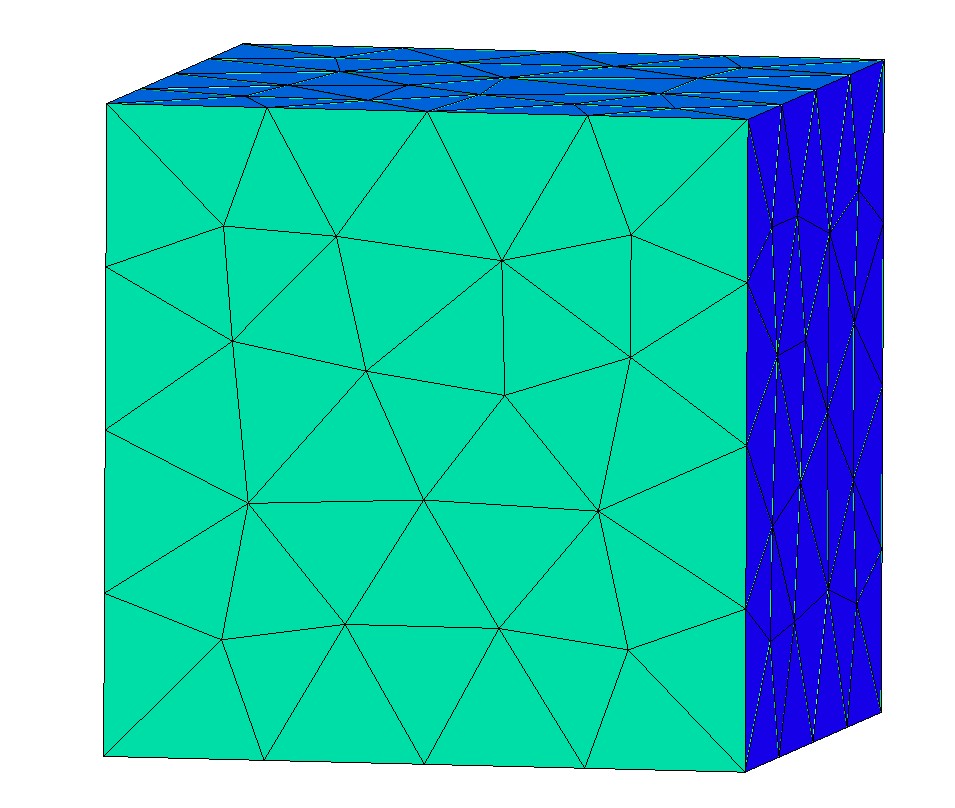}
		\caption{Coarse representation}
		\label{Coarserep}
	\end{subfigure}
	\hfill
	\begin{subfigure}{.35\textwidth}
		\centering
		\includegraphics[width=0.9\textwidth]{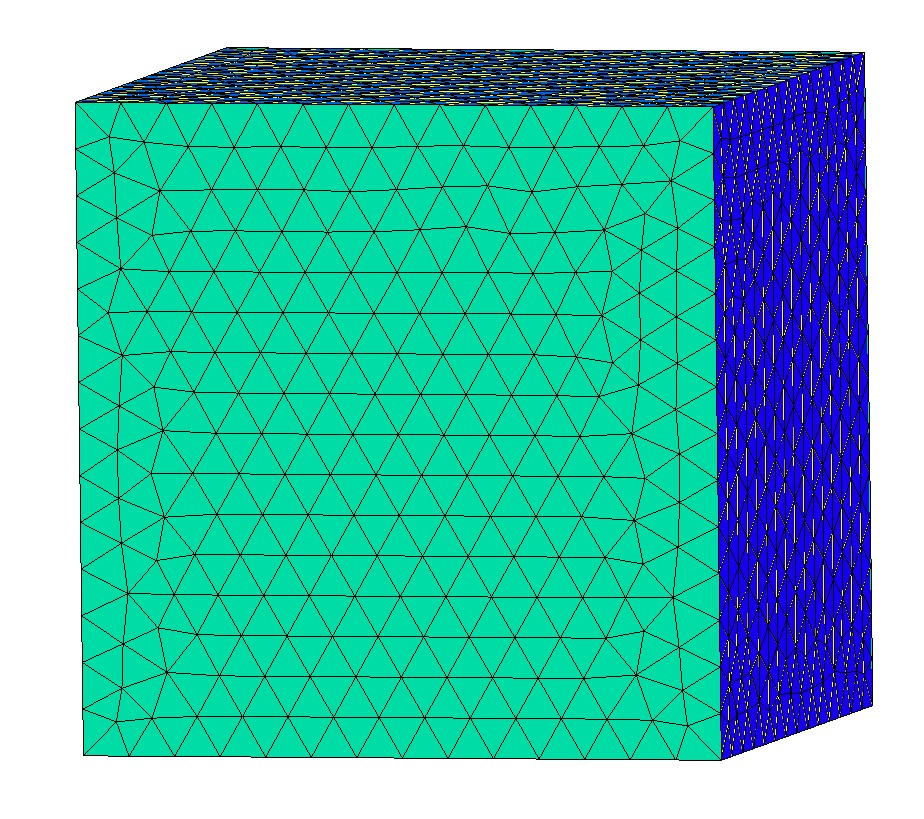}	
		\caption{Fine representation}
		\label{Finrep}
	\end{subfigure}
	\hfill
	\begin{subfigure}{.2\textwidth}
		\centering
		\includegraphics[width=0.9\textwidth]{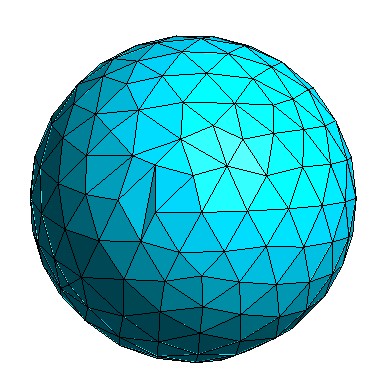}	
		\caption{Spherical inclusion}
		\label{Spherep}
	\end{subfigure}\hfill\null
	\caption{Example of one unit cube with different representations}
\end{figure}

Two representations of cubes are considered in the study. The cubes in the global model are homogenous and have a coarse mesh, as depicted in Figure~\ref{Coarserep}. On the other hand, the cubes in the local models have refined meshes (shown in Figure~\ref{Finrep}). They are heterogeneous and contain a spherical inclusion within each cube. By default, the sphere is centered and has a radius equal to a quarter of the side of the cube, although these parameters may vary in certain studies (refer to Figure~\ref{Spherep}). It should be noted that the fine meshes are independently constructed within the patches and are not required to match at the interface. In contrast, the global mesh conforms to the interface.


There are two linear problems to consider: the Poisson equation, which is used to model thermal problems, and the linear elasticity equation. In the case of the Poisson equation, it represents the conduction of heat and assumes homogeneous boundary conditions. In some scenarios, there may be a contrast in the conductivity coefficient, denoted as "a". The source term in this equation is simply set to 1. The linear elasticity equation, on the other hand, is used to model mechanical deformation. Here, a contrast in Young's modulus is introduced. The source term is a vector with all components equal to 1, representing a uniform force or load applied to the system.
%

\subsection{Rate of communication study}
To gain a better understanding of the impact of synchronization on communication and waiting time, a first study is proposed. The study focuses on a thermal problem set on a well-balanced case with 64 subdomains. The case has a heterogeneity ratio of 100, where the thermal diffusion coefficient in the spherical inclusion is 100 times lower than in the rest of the cube. The dimensions of the problems are provided in Table~\ref{64_mesh}.

Different numbers of MPI processes are employed, which means that for fewer than 65 processes, a single MPI rank is responsible for handling multiple subdomains. The study analyzes the percentage of communication time in the total simulation time, considering a convergence tolerance of $10^{-7}$ for the norm of the residual.

\begin{table}[H]
	\centering
	\begin{tabular}{|l|c|c|}
		\hline
		Problem&  \textbf{Global} & \textbf{Local (One subdomain)} \\ 
		Number of nodes & 1449  &  1858 \\ \hline
	\end{tabular}
	\caption{Mesh data}
	\label{64_mesh}
\end{table}

\begin{table}[ht]\centering
	\begin{tabular}{|l||c|c|}\hline
		\textbf{\#ranks}  	&\textbf{Aitken}        	&\textbf{Asynchronous}  \\ 
		&\#iter 					&\#iter. glob.[\#loc. sol. [min, max]] \\
		&\#time (s)         		&\#time (s) \\ 
		&[\% communication time]    & [\%  communication time]         \\  \hline
		\textbf{9}     		&25 \& 11.72s [30\%]        &  334[49, 54] \& 22.4s[10\%]     \\  \hline
		\textbf{17}    		&25 \& 8.08s  [80\%]        &  182[56, 77] \& 13.25s[10\%]    \\  \hline
		\textbf{33}      	&25 \& 4.53s [71\%]         &  104[65, 124]\& 8.13s[16\%]     \\  \hline
		\textbf{65}   	    &25 \& \bf{8.57s [97\%]}    &  105[81, 160] \& \bf{8.40s[46\%]}\\  \hline
	\end{tabular}
	\caption{Analysis of the time spent in communication (64-subdomain thermal case).}
	\label{ratecom}
\end{table}

Table~\ref{ratecom} illustrates that the synchronous approach with the Aitken accelerator outperforms the asynchronous approach in terms of overall speed. However, it is noteworthy that the proportion of time dedicated to communication significantly increases in the synchronous case, reaching up to 97\%. In contrast, the asynchronous case experiences a more moderate increase, never exceeding 50\%. As a result, the asynchronous approach proves to be faster in the 65-process case. Specifically, the transition from computations involving a single node (9 subdomains) to two nodes (17 subdomains) leads to a substantial increase in communication time for the synchronous case, while the asynchronous case remains unaffected.

\subsection{Load imbalance study}
In contrast to the previous studies that focused on nearly perfect load balance, this section examines the case of load imbalance to demonstrate the impact of synchronization and highlight the advantages of the asynchronous model.

\begin{figure}[ht]
	\centering
	\includegraphics[width=0.5\textwidth]{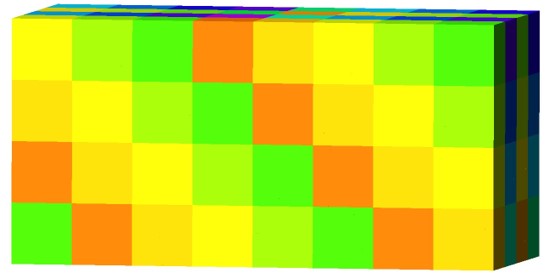}
	\caption{128 subdomains with $\Omega^{0} = \emptyset$ }
	\label{128}
\end{figure}

The study is conducted using a 128-patch geometry (Figure~\ref{128}) with dimensions provided in Table~\ref{128_mesh}. The Young's modulus in the spherical inclusion is 100 times lower than in the remaining part of the cube. The objective is to achieve a global residual norm below $10^{-7}$.
\begin{table}[ht]
	\centering
	\begin{tabular}{|l|c|c|}
		\hline
		Problem&  \textbf{Global} & \textbf{Local (One subdomain)} \\ \hline
		\hline
		Numbre of nodes & 2769  &  1254 \\ \hline
	\end{tabular}
	\caption{Mesh data}
	\label{128_mesh}
\end{table}

In this scenario, the objective is to assign varying computational loads to each core, assuming a limited number of cores available, specifically 65 cores. The task distribution is random, and two cases are considered:
\begin{enumerate}
	\item  Access to all machines in the cluster is granted, but only a specific number of cores are allocated per machine. As a result, the machines operate at a reduced capacity. However, the exchange of information becomes more extensive as it involves multiple machines.
	
		\begin{tabular}{|l|c|c|}
			\hline
			Model&  \textbf{Aitken} & \textbf{Relaxed asynchronous} \\ \hline
			\hline
			Time(s) & 150  &  130.07 \\ \hline
			Async. \#iter. glob.  &   24 & 239 \\ \hline
			Async. \#loc. sol. [min, max] &   - & [70 - 338]  \\ \hline
		\end{tabular}
	
	In the asynchronous case, despite a large number of iterations performed both globally and locally, it still exhibits faster performance. The noticeable discrepancy between the minimum and maximum number of iterations, caused by load imbalance, is evident. This condition, which poses a disadvantage for the synchronous approach, highlights the adaptability of the asynchronous method to load imbalance situations.
	
	\item Considering a limited number of machines that can accommodate up to 65 cores, the focus is on utilizing the machines intensively while keeping the exchange network less burdened.
	
		\begin{tabular}{|l|c|c|}
			\hline
			Model&  \textbf{Aitken} & \textbf{Relaxed asynchronous} \\ \hline
			\hline
			Time(s) & 286.9  & 	177.27s \\ \hline
			Async. \#iter. glob.   &   24 & 353  \\ \hline
			Async. \#loc. sol. [min, max] &   - & [69 - 417]  \\ \hline
		\end{tabular}
	
	In this scenario, the overall computation times are longer compared to the previous situation. The high demand placed on the machines adversely affects the synchronous algorithm, resulting in slower performance. On the other hand, the asynchronous algorithm demonstrates significantly faster execution in this context.
	
\end{enumerate}
\section{conclusion}
We have introduced an asynchronous version of the non-intrusive global/local computation method and proposed its implementation using MPI RMA parallelization techniques. The global/local framework is advantageous for asynchronism as we can easily prove and control its convergence.

The presented performance results are promising: in terms of computation time, the asynchronous method approaches the formidable Aitken variant of synchronous computation.
\section*{Acknowledgements}
This work was partly funded by the French National Research Agency as part of project ADOM, under grant number ANR-18-CE46-0008.


\begin{thebibliography}{99}
	\bibitem{cormier_Aggressive_1999}
	Nathan Cormier, Brian S. Smallwood, Gleen B. Sinclair, and G. Meda. "Aggressive
	submodelling of stress concentrations". International Journal for Numerical Methods in Engineering, 46(6):889–909, 1999.
	\bibitem{mpi4py}
	Lisandro Dalcin and Yao-Lung L. Fang. "mpi4py: Status update after 12 years of development". Computing in Science \& Engineering, 23(4):47–54,
	2021.
	\bibitem{duval2014non}
	Mickaël Duval, Jean-Charles Passieux, Michel Salaün, and St\'ephane
	Guinard. "Non-intrusive coupling: recent advances and scalable nonlinear
	domain decomposition". Archives of Computational Methods in Engineering,
	23(1):17–38, 2014.
	\bibitem{ELKERIM2023115910}
	Ahmed El Kerim, Pierre Gosselet, and Fr\'ed\'eric Magoul\`es. "Asynchronous
	global-local non-invasive coupling for linear elliptic problems". Computer
	Methods in Applied Mechanics and Engineering, 406:115910, 2023.
	\bibitem{Garay2022SynchronousAA}
	Jos\'e C. Garay, Fr\'ed\'eric Magoul\`es, and Daniel B. Szyld. "Synchronous and
	asynchronous optimized Schwarz methods for Poisson’s equation in rectangular
	domains". ETNA - Electronic Transactions on Numerical Analysis, 2022.
	\bibitem{Gbikpi-Benissan2022}
	Guillaume Gbikpi-Benissan and Fr\'ed\'eric Magoul\`es. "Resilient asynchronous
	primal schur method". Applications of Mathematics, 67:679-704, 2022.
	\bibitem{gendre2009non} 
	Lionel Gendre, Olivier Allix, Pierre Gosselet, and François Comte. "Nonintrusive
	and exact global/local techniques for structural problems with
	local plasticity". Computational Mechanics, 44(2):233–245, 2009.
	\bibitem{GMSH}
	Christophe Geuzaine and Jean-François Remacle. "gmsh : a threedimensional
	nite element mesh generator with built-in pre- and postprocessing
	facilities". International Journal for Numerical Methods in Engineering,
	0:1–24, 2009.
	\bibitem{Glusa}
	Christian Glusa, Erik G. Boman, Edmond Chow, Sivasankaran Rajamanickam,
	and Daniel B. Szyld. "Sacalable asynchronous domain decomposition
	solvers". SIAM Journal on Scientific Computing, 42(6):384–409, 2020.
	\bibitem{blanchard.2018.1}
	Pierre Gosselet, Maxime Blanchard, Olivier Allix, and Guillaume Guguin.
	"Non-invasive global-local coupling as a Schwarz domain decomposition
	method: acceleration and generalization". Advanced Modeling and Simulation
	in Engineering Sciences, 5(4), 2018.
	\bibitem{hecht_2009_nzs}
	Fr\'ed\'eric Hecht, Alexei Lozinski, and Olivier Pironneau. Numerical zoom
	and the Schwarz algorithm. In Proceedings of the 18th conference on domain
	decomposition methods, 2009.
	\bibitem{kelley1982}
	FS Kelley. "Mesh requirements for the analysis of a stress concentration by
	the specified boundary displacement method". In Proceedings of the Second
	International Computers in Engineering Conference, ASME, pages 39–42, 1982.
	\bibitem{magoules2018asynchronous}
	Fr\'ed\'eric Magoul\`es and C\'edric Venet. "Asynchronous iterative substructuring
	methods". Mathematics and Computers in Simulation, 145:34-49,
	2018.
	\bibitem{GetFEM}
	Yves Renard and Konstantinos Poulios. "Getfem: Automated fe modeling
	of multiphysics problems based on a generic weak form language". Advances
	in Engineering Software, 47:1–31, 2021.
	\bibitem{Yamazaki.19}
	Ichitaro Yamazaki, Edmond Chow, Aurelien Bouteiller, and Jack Dongarra.
	"Performance of asynchronous optimized schwarz with one-sided communication".
	Parallel Computing, 86:66 – 81, 2019.
	\bibitem{ransom1992computational}
	Ransom, Jonathan B and McCleary, Susan L and Aminpour, Mohammad A and Knight Jr, Norman F. 
	"Computational methods for global/local analysis". NASA STI/Recon Technical Report N 92, 33104, 1992.	
\end{thebibliography}
\end{document}